\documentclass[10pt]{article}
\usepackage{longtable}
\usepackage{bbm}

\usepackage{maple2e}

\DefineParaStyle{Maple Output}
\DefineCharStyle{2D Math}
\DefineCharStyle{2D Output}
\DefineParaStyle{Bullet Item}
\DefineParaStyle{Heading 1}
\DefineParaStyle{Heading 2}
\DefineParaStyle{Warning}
\DefineCharStyle{2D Comment}
\DefineCharStyle{Help Normal}
\DefineCharStyle{Hyperlink}

\usepackage{amssymb}
\usepackage{amsmath}
\usepackage{algorithmic}
\usepackage{graphicx}
\usepackage{amsthm}
\newcommand{\Gr}{Gr\"obner }

\newcommand{\Q}{\mathbb{Q}}

\newcommand{\N}{\mathbb{N}}

\newcommand{\R}{\mathbb{R}}

\newcommand{\lm}{\mathop{\mathrm{lm}}\nolimits}
\newcommand{\lt}{\mathop{\mathrm{lt}}\nolimits}
\newcommand{\lc}{\mathop{\mathrm{lc}}\nolimits}

\def \bg #1 {\begin{tabular}{{#1}}}
\def \nd {\end{tabular}}

\newenvironment{algorithm}[1]{
  \begin{center}
    {\bf Algorithm: #1}\\*
     \begin{tabular}{|p{130mm}|} \hline
} {
 \\ \hline
 \end{tabular}
 \end{center}
}

\begin{document}

\title{\bf \Gr Bases Applied to Systems of Linear Difference Equations
}

\author{Vladimir P. Gerdt \\
       Laboratory of Information Technologies\\
       Joint Institute for Nuclear Research\\
       141980 Dubna, Russia \\
       }
\date{}
\maketitle

\begin{abstract}
In this paper we consider systems of partial (multidimensional) linear difference equations. Specifically, such systems
arise in scientific computing under discretization of linear partial differential equations and in computational
high energy physics as recurrence relations for multiloop Feynman integrals. The most universal algorithmic tool for
investigation of linear difference systems is based on their transformation into an equivalent \Gr basis form. We
present an algorithm for this transformation implemented in Maple. The algorithm and its implementation can be applied
to automatic generation of difference schemes for linear partial differential equations and to reduction of Feynman
integrals. Some illustrative examples are given.
\end{abstract}

\section{Introduction}

Let $\N_{>0}$ and $\N_{\geq 0}$ be the sets of positive and nonnegative integers, $\Q$ be the set of
rational numbers, $Y:=\{\,y^j(x_1,\ldots,x_n)\}\mid j=1,\ldots,m\,\ m,n\in \N_{>0}\}$ be the set of
functions in $n$-variables, and $\theta_i$ be the right-shift operator for the $i$-th variable:
$
\theta_i\circ y(x_1,\ldots,x_n):=y(x_1,\ldots,x_i+1,\ldots,x_n)\,.
$
For the power products $\theta_1^{\mu_1}\cdots \theta_n^{\mu_n}$ of the shift operators we shall use the multiindex
notation $\theta^\mu$ where $\mu:=\{\mu_1,\ldots,\mu_n\}$ ($\mu\in \N_{\geq 0}^n$) with $\mu:=\sum \mu_i$. The set of all such operator products
will be denoted by $\Theta$.

And then the most general form of a system of $K\in \N_{> 0}$ partial $(n > 1)$ and multivariate $(m > 1)$ linear
difference equations is given by
\begin{equation}\label{system}
a_0+\sum_{j=1}^m \sum_{\nu} a_{kj;\,\nu}\vartheta^\nu_k \circ y^j=0\,,\qquad k=1,\ldots,K\,,\qquad \vartheta_k^\nu\in \Theta\,,
\end{equation}
where all sums are finite and coefficients $a_0,a_{kj;\,\nu}$ may depend on the variables $X:=\{x_1,\ldots,x_n\}$ and on
a finite set of parameters $C:=\{c_1,\ldots\}$. Hereafter we shall assume that all coefficients in~(\ref{system})
are rational functions of the variables and parameters with integer coefficients:
\begin{equation}\label{coeff}
a_0,a_{kj;\,\nu}\in \Q(X\cup C)\,.
\end{equation}
This restriction on the coefficients allows to apply algorithmic technique of the
next section.

It is well-known that, except very simple cases, systems of form~(\ref{system}-\ref{coeff}) do not admit exact
solutions and rather weakly
studied in the literature~\cite{KP'00}. However, such systems play a fundamental role in a number of important
applications for instance in:
\begin{description}
\item[Scientific Computing:] Numerical solving of linear partial differential systems~\cite{H'68} with rational function
 coefficients. Recently, it was observed~\cite{GBM'06} that one can automatically generate finite-difference schemes
 for such partial differential equations (PDEs) by eliminating partial derivatives from certain linear partial
 and multivariate difference systems. In so doing, for homogeneous PDEs whose coefficients may also be rational
 functions of parameters one deals with systems of form~(\ref{system}-\ref{coeff}) with $a_0=0$.
\item[Computational High Energy Physics:] Reduction of multiloop Feynman integrals~\cite{Smirnov'04}. These integrals, after a proper
 fixed right shift of the variables in $X$ satisfy the univariate system of partial difference equations
 (recurrence relations)~\cite{ChTk'81} whose rational function coefficients depend on such physical parameters as
 the space-time dimension, masses and external momenta. The problem is to reduce the integrals to be
 evaluated to a minimal set of basic or master integrals, i.e., those integrals which are independent modulo the
 difference system, and then to express other integrals in terms of the basic ones.
\item[Computational Economics:] Characterization of economic behavior in macroeconomics \cite{WW'06}. Here for some
 macroeconomic problems one has to solve system~(\ref{system}) with constant parametric coefficients.
\end{description}

To investigate or to solve difference systems (\ref{system}) with rational coefficients~(\ref{coeff}) one can use the
universal algorithmic \Gr bases method invented about 40 years ago by Buchberger~\cite{Buch'65} for systems of
multivariate commutative polynomials generating polynomial equations~\cite{Buch'85}. The main idea of this method is to
rewrite the initial system of equations into a certain equivalent form called a \Gr basis which makes easier
investigation of the system and its solving. The underlying Buchberger's algorithm~\cite{Buch'85} built-in all modern
general-purpose computer algebra systems such as Maple, Mathematica and others.

On the basis of research made to date, the \Gr bases theory was extended to some ``weakly'' noncommutative
polynomials as well as to linear differential or difference polynomials and operators~\cite{BW'98,KLMP'99}.
Generally, however, the noncommutative and nonlinear differential or difference \Gr bases may not exist (be
infinite). For difference systems~(\ref{system}-\ref{coeff}) \Gr base are always finite and can be constructed by
Buchberger's algorithm straightforwardly translated to difference algebra~\cite{KLMP'99}.

Recently~\cite{GBM'06,G'06} we presented the difference form of our polynomial algorithm devised
in~\cite{GB'98}, improved in~\cite{G'05} and specialized to so-called Janet and Janet-like monomial divisions~\cite{GB'05} which
go back to the constructive ideas of French mathematician Janet~\cite{Janet'29}. The algorithm
constructs a Janet(-like) basis~\cite{GB'05} which is also a \Gr basis. Though generally Janet bases~\cite{GB'98} and
Janet-like bases are redundant as \Gr ones, the algorithm in its improved version~\cite{GBM'06,G'05} allows also to
output reduced \Gr bases without any additional computational costs. The implementation~\cite{GR'06} of the algorithm in
Maple allows a user to compute linear difference Janet(-like Gr\"{o}bner) bases.

In the present paper we describe briefly a simple version of the Janet division algorithm (Sect.2) and consider its application to
the above listed problems from scientific computing (Sect.3) and computational high energy physics (Sect.4). Both problems are purely
algebraic and can be completely solved with the use of \Gr bases. We illustrate this fact by simple examples. Our presentation is
addressed to non-algebraists. By this reason we slightly abuse algebraic terminology and refer to the references in
bibliography for more precious definitions and notions.

\section{Transformation to \Gr Basis}
In this section we define the concept of a \Gr basis form for the difference system~(\ref{system}-\ref{coeff}) and present an
algorithm for its computation. The \Gr basis form of system~(\ref{system}) is defined by a ranking $\succ$
(linear order on) of terms $\theta^\mu \circ y^{\,j}$ and such that for all $i,j,k,\mu,\nu$ the following holds:
\begin{displaymath}
\theta_{i} {\theta^\mu \circ y^{\,j}} \succ {\theta^\mu}\circ y^{\,j}\,,\qquad
\theta^\mu \circ y^{\,j} \succ \theta^\nu \circ y^k \iff
{\theta_i}  {\theta^\mu \circ y^{\,j}}
   \succ {\theta_i}{\theta^\nu} \circ y^k\,.
\end{displaymath}
If $|\mu| \succ |\nu| \Longrightarrow {\theta^\mu \circ y^{\,j}} \succ {\theta^\nu} \circ y^k$
the ranking is called {\it orderly}. If $j > k \Longrightarrow {\theta^\mu} \circ y^{\,j} \succ {\theta^\nu} \circ y^k$
the ranking is called {\it elimination}.

Denote by $f_k$ the left-hand side of the $k$-th equation in~(\ref{system}) and by $F:=\{f_1,\ldots,f_K\}$ the set of all
the left-hand sides in the system. Fixing a ranking $\succ$ provides every $f\in F$ with the {\it leading term}
$\lt(f):=a\,\vartheta \circ y^j$ ($\vartheta\in \Theta,\,a\neq 0$) and {\it leading coefficient} $\lc(f):=a$.
Furthermore, denote $R\supset F$ the set of all right-hand sides $f\neq 0$ for linear difference equations $f=0$ which are consequences
of system~(\ref{system}-\ref{coeff}). $F$ is called {\it generating set or basis} of $R$ (denotation: $R=<F>$). In that follows we shall assume that,
given a ranking $\succ$, all $f\in R$ are normalized, that is, divided by their leading coefficients.
If $F \subseteq \R$, then $\lt(F)$ will denote the set of the leading terms and $\lt_j(F)$ will denote
its subset for function $y^{\,j}$. Therefore,
\begin{displaymath}
\lt(F)=\cup_{j=1}^m \lt_j(F)\,.
\end{displaymath}

\noindent
Now we are ready to define a {\it \Gr basis} for given $F$ and ranking $\succ$ as a finite subset $G\subset R=<F>$ such that
$R=<G>$ and
\begin{equation}
\forall f\in R\,,\ \exists \, g\in G,\, \theta \in \Theta\  :\
\lt(f)=\theta \circ \lm(g)\,. \label{GB}
\end{equation}
It follows that the leading term of every $f\in R$ is
{\it reducible modulo $G$} and yields {\it the head reduction}:
\begin{displaymath}
f \xrightarrow[g]{} f':=f-\theta \circ g,\quad f'\in R\,.
\end{displaymath}
If $f'\neq 0$, then its leading term is again reducible modulo $G$. And
then by repeating the reduction finitely many times~\cite{Buch'85,BW'98,KLMP'99} we obtain
$ f \xrightarrow[G]{}0$.
Generally, if a linear difference expression $h$ (not necessarily from $R$) contains a term $u$ with coefficient $c\neq 0$
such that $u=c\,\vartheta \circ \lt(f)$ for some $\vartheta \in \Theta$ and
$f\in F\subset \R$, then $h$ can be reduced:
\begin{equation}
h \xrightarrow[g]{} h':=h-c\,\vartheta \circ f\,. \label{elem_red}
\end{equation}
By applying the reduction finitely many times, one obtains a polynomial $\bar{h}$ which is either zero or
such that all its (nonzero) terms are {\it irreducible modulo set $F$}.
In both cases $\bar{h}$ is said to be in the {\it normal form modulo $F$}
(denotation: $\bar{h}=NF(h,F)$). A \Gr basis $G$ is called {\it reduced} if
$g=NF(g,G\setminus \{g\})$ for every $g\in G$.

In our algorithmic construction of reduced \Gr bases we shall use a restricted
set of reductions called {\it Janet reductions} (cf.~\cite{G'05}) and defined
as follows.

For a finite set $F$ and a ranking $\succ$, we partition
every set $\lt_k(F)$ into groups
labeled by $d_0,\ldots,d_i\in \N_{\geq 0}$,\ $(0\leq i\leq n)$. Here $[0]_k:=\lt_k(F)$
and for $i>0$ the group $[d_0,\ldots,d_i]_k$ is defined as
\begin{displaymath}
[d_0,\ldots,d_i]_k:=\{u\in \lt_k(F) \mid
d_0=0,d_j=\deg_j(u),1\leq j\leq i \}
\end{displaymath}
where $\deg_i(\theta^\mu\circ y^k):=\mu_i$. Operator $\theta_i$ is called {\em $J$(anet)-multiplicative} for $f\in F$ if
$\lt(f)\in [d_0,\ldots,d_{i-1}]$ and $\deg_i(u)=\max\{\deg_i(v) \mid v\in [d_0,\ldots,d_{i-1}]\}$.
Denote by $M_J(f,F)$ the set of {\em $J$(anet)-multiplicative} shift operators for $f\in F$, the complement
set $\{\theta_1,\ldots,\theta_n\}\setminus M_J(f,F)$ of {\it $J$(anet)-nonmultiplicative} shift operators by $NM_J(f,F)$
and the set of all possible power products of $J-$multiplicative operators (including identity operator) by $J(f,F)$.
It is clear that $J(f,F)\subset \Theta $.

A finite set $G\in R=<F>$ is called a {\it Janet basis}~(cf.\cite{G'05}) if
\begin{equation}
\forall f\in R\,,\exists \, g\in G, \theta
\in {J}(g,G)\  :\ \lt(f)=\theta \circ
  \lt(g)\,. \label{JB}
\end{equation}
Similarly to~(\ref{elem_red}), a {\em ${J}-$reduction} is defined as
\begin{equation}
h \xrightarrow[g]{} h':=h-c\,\vartheta \circ f\,,\quad \vartheta\in {J}(f,F)\,, \label{elem_J_red}
\end{equation}
for a polynomial $h\in R$ containing a term $u$ with coefficient $c\neq 0$
satisfying $u=c\,\vartheta \circ \lt(f)$ for some $f\in F$  and
$\vartheta \in {J}(f,F)$.

Since $J-$reducibility~(\ref{elem_J_red}) implies the \Gr reducibility (\ref{elem_red}), a Janet basis
satisfying~(\ref{JB}) is also a \Gr basis. The converse is generally not
true, that is, not every \Gr basis is Janet one. The algorithmic characterization of a Janet basis $G$ is
the following condition~(cf.~\cite{G'05}):
\begin{equation}
\forall g\in G,\ \theta\in NM(g,G):\,NF_{J}(\theta \circ g,G)=0\,. \label{alg_char}
\end{equation}
which is a cornerstone of the below algorithm for construction of Janet bases~(\ref{GB}).

This algorithm is a translation (with some minor modifications) of the polynomial algorithm in~\cite{GB'06} into the
difference case. Due to the normalization of $h$ done at Step.15 before insertion of $h$ into the intermediate basis $G$,
the algorithm outputs the minimal and normalized Janet basis which is uniquely defined by an input difference system
$F$ and a ranking~\cite{GB'98}. Correctness and termination of the difference algorithm immediately follow from those
for its polynomial counterpart~~\cite{GB'98,G'05}. Algorithm {\bf JanetBasis} implemented in its improved
form~\cite{GBM'06} as the Maple package LDA
(abbreviates {\underline{L}inear \underline{D}ifference \underline{A}lgebra)~\cite{GR'06}, and in the next
two sections computation with the package is illustrated by examples.

\begin{algorithm}{JanetBasis($F,\succ$)\label{JanetBasis}}
\begin{algorithmic}[1]
\INPUT $F$, a finite linear difference set;\ $\succ$, a ranking
\OUTPUT $G$, a Janet basis of $<F>$
\STATE {\bf choose} $f\in F$ with the lowest $\lt(f)$ w.r.t. $\succ$
\STATE $G:=\{f\}$
\STATE $Q:=F\setminus G$
\DOWHILE
  \STATE $h:=0$
  \WHILE{$Q\neq \emptyset$\ and $h=0$}
    \STATE {\bf choose} $p\in Q$ with the lowest $\lt(p)$ w.r.t. $\succ$
    \STATE $Q:=Q\setminus \{p\}$
    \STATE $h:=NF_J(p,G)$
  \ENDWHILE
  \IF{$h\neq 0$}
    \FORALL{$g\in G$ such that $\lt(g)=\theta^\mu \circ \lt(h),\ |\mu|>0$}
      \STATE $Q:=Q\cup \{g\}$; \ $G:=G\setminus \{g\}$
    \ENDFOR
    \STATE $G:=G\cup \{ h/\lc(h) \}$
    \STATE $Q:=Q\cup \{\,\theta^\beta\circ g \mid g\in G,\
                                    \theta^\beta\in NM_J(g,G)\,\}$
 \ENDIF
\ENDDO{$Q \neq \emptyset$}
\RETURN $G$
\end{algorithmic}
\end{algorithm}

\section{Generation of Difference Schemes}

In paper~\cite{GBM'06} an algorithmic approach was developed to construct finite-difference schemes for linear
PDEs in two independent variables and on uniform orthogonal grids with possibly distinct mesh steps for $x$ and $y$.
We outline here the main idea of the approach and refer to~\cite{GBM'06} for more details. In so doing, we restrict
our consideration by scalar equations of order $\geq 2$ which admit the conservation law form
\begin{equation}
\frac{\partial V}{\partial x} + \frac{\partial W}{\partial y}=0 \label{cons_law}
\end{equation}
where $V$ and $W$ are functions of independent variables $x,y$, dependent variable $u(x,y)$ and its partial
derivatives $u_x,u_y$, ${u}_{xx},\ldots$. Differential equation~(\ref{cons_law}) can be rewritten
in the integral form
\begin{equation}
 \oint \limits_{\Gamma} \!- Wdx + Vdy = 0\, \label{int_cons_law}
\end{equation}
which is valid for arbitrary closed contour $\Gamma$. Discretization of~(\ref{int_cons_law}) instead of~(\ref{cons_law})
is natural for preserving the conservation low at the discrete level (conservative scheme).

Denote the grid values of function $u(x, y)$ and its derivatives by
\begin{equation}
u_{j\,k}:=u(x_j,y_k),\ (u_x)_{j\,k}:=u_x(x_j,y_k),\ (u_y)_{j\,k}:=u_y(x_j,y_k),\ (u_{xx})_{j\,k}:=u_{xx}(x_j,y_k),\
\ldots\ ,\label{grid}
\end{equation}
and fix some integration contour $\Gamma$ in (\ref{int_cons_law}) on the grid. To be specific, let us choose the
following simple rectangular contour
{\footnotesize
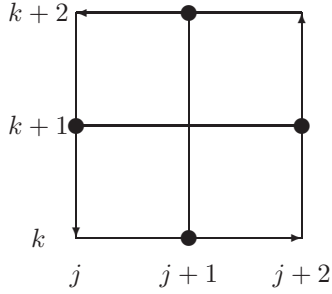
\begin{figure}[htbp]
\unitlength=1.00mm
\special{em:linewidth 0.4pt}
\linethickness{0.4pt}
\begin{center}
\begin{picture}(110.00,41.00)(-18, 0)
\put(21.00,10.00){\vector(1,0){30.00}}
\put(51.00,10.00){\vector(0,1){30.00}}
\put(51.00,40.00){\vector(-1,0){30.00}}
\put(21.00,40.00){\vector(0,-1){30.00}}
\put(21.00,25.00){\line(1,0){30.00}}
\put(36.00,10.00){\line(0,1){30.00}}
\put(36.00,40.00){\circle*{2.00}}
\put(51.00,25.00){\circle*{2.00}}
\put(21.00,25.00){\circle*{2.00}}
\put(36.00,10.00){\circle*{2.00}}
\put(16.00,10.00){\makebox(0,0)[cc]{$k$}}
\put(16.00,25.00){\makebox(0,0)[cc]{$k+1$}}
\put(16.00,40.00){\makebox(0,0)[cc]{$k+2$}}
\put(21.00,5.00){\makebox(0,0)[cc]{$j$}}
\put(36.00,5.00){\makebox(0,0)[cc]{$j+1$}}
\put(51.00,5.00){\makebox(0,0)[cc]{$j+2$}}
\end{picture}
\caption[Integration contour on grid]{
   \label{int_cont}
   Integration contour on grid}
\end{center}
\end{figure}
}

Now we add to the integral equation (\ref{int_cons_law}) for the rectangular contour of Fig.~\ref{int_cont}
all the related (exact) integral relations between $u(x,y)$ and its partial derivatives:
\begin{equation}
\left\{
\begin{array}{l}
\int \limits_{x_j}^{x_{j+2}} \! \! \! {u}_x dx = {u}(x_{j+2}, y) -
{u}(x_{j}, y)\,,\quad
\int \limits_{y_k}^{y_{k+2}} \! \! \! {u}_y dy = {u}(x, y_{k+2}) -
{u}(x, y_{k})\,, \\
 \int \limits_{x_j}^{x_{j+2}} \! \! \! {u}_{xx} dx = {u}_x(x_{j+2}, y) -
{u}_x(x_{j}, y)\,,\quad
\int \limits_{y_k}^{y_{k+2}} \! \! \! {u}_{xy} dy = {u}_x(x, y_{k+2}) -
{u}_x(x, y_{k})\,,  \\
.....................................................................................................................
\end{array}
\right.
\label{rel}
\end{equation}

\noindent
Our purpose is to obtain a difference scheme for $u_{j\,k}$ from a proper discretization of integral equations
(\ref{int_cons_law}) and relations~(\ref{rel}). To do that one should use as many relations in~(\ref{rel})
as the number of all proper derivatives of $u$ up to the maximal orders of their occurrence in the integrand
of~(\ref{int_cons_law}). Then the difference scheme can be obtained by {\it an algebraic difference elimination} of all
discrete proper partial derivatives in list~(\ref{grid}) from the combined system~(\ref{int_cons_law},\ref{rel}).
The algebraic elimination can be achieved by computing a \Gr or Janet basis for the last system and a suitable
elimination ranking (see Sect.3) satisfying $u_{j\,k}\prec (u_x)_{j\,k}\prec (u_{xx})_{j\,k}\prec \cdots$.

Therefore, to construct an initial system of discrete equations for the following difference elimination, it suffices
to approximate numerically the contour integral~(\ref{int_cons_law}) for the chosen contour
of Fig.~\ref{int_cont} together with the integral relations~(\ref{rel}) in terms of the
grid unknowns~(\ref{grid}). For this purpose one can choose various quadrature formulas for these integrals, and
the difference scheme obtained may depend on the choice. For simplicity sake we apply here for all the integrals in
(\ref{int_cons_law}) and \ref{rel}) the simplest rectangle (midpoint) rule:
\begin{equation}
  \left\{
  \begin{array}{l}
  ({W}_{j+1 \, k+2}  -{W}_{j+1 \, k})\cdot h_1 +
  ({V}_{j+2 \, k+1} - {V}_{j \, k+1})\cdot h_2   = 0\,, \\
 ({u}_x)_{j+1 \, k} \cdot 2 h_1 = {u}_{j+2 \, k} - {u}_{j \, k}\,, \\
 ({u}_y)_{j \, k+1} \cdot 2 h_2 = {u}_{j \, k+2} - {u}_{j \, k}\,,\\
 ................................................
 \end{array}
 \right.
 \label{disc_system}
\end{equation}
where $h_1:=x_{j+1} -  x_{j}$ and  $h_2:= y_{k+1} -  y_{k}$ are the grid mesh steps for our uniform orthogonal grid.

For linear difference system~(\ref{disc_system}) Janet ( \Gr) basis exists for any ranking, and, hence, the elimination
can be performed by applying the above algorithm {\bf JanetBasis}. To illustrate this algorithmic procedure for the
difference schemes generation consider a simple example of the Heat equation in its conservation law form~\cite{GBM'06}:
\begin{equation}
u_{t} + \alpha u_{xx}=0\quad \Longrightarrow\quad  \oint \limits_{\Gamma} \!-  \alpha u_x dt + u dx = 0\,. \label{int_HE}
\end{equation}
where $\alpha$ is a symbolic parameter. The integrand in~(\ref{int_HE}) contains the only partial derivative $u_x$.
Hence, we need to add the only integral relation
\begin{equation}
 \int \limits_{x_j}^{x_{j+1}} \! \! \! u_x dx = u(x_{j+1}, t) - u(x_{j}, t)\,. \label{IR_for_HE}
\end{equation}

Now consider $u(x,t)$ and $u_x(x,t)$ on the uniform orthogonal grid with the spatial mesh step $h$
and the temporal mesh step $\tau$, and choose the simplest contour shown in~Fig.~\ref{HE_int_cont}. As this takes place,
we can approximate the integral of $u_x(x,t)$ over $x$ in~(\ref{int_HE}-\ref{IR_for_HE}) on the grid points by the
rectangular or trapezoidal rules.

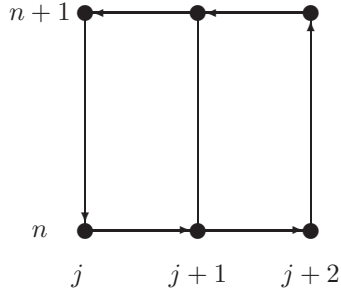
\begin{figure}
\begin{center}
\unitlength=1.00mm
\special{em:linewidth 0.4pt}
\linethickness{0.4pt}
\begin{picture}(110.00,39.00)(40, 4)
\put(79.00,40.00){\vector(0,-1){28.00}}
\put(79.00,40.00){\circle*{2.00}}
\put(109.00,40.00){\circle*{2.00}}
\put(109.00,11.00){\circle*{2.00}}
\put(79.00,11.00){\circle*{2.00}}
\put(109.00,11.00){\vector(0,1){28.00}}
\put(73.00,11.00){\makebox(0,0)[cc]{$n$}}
\put(73.00,40.00){\makebox(0,0)[cc]{$n+1$}}
\put(78.00,5.00){\makebox(0,0)[cc]{$j$}}
\put(109.00,5.00){\makebox(0,0)[cc]{$j+2$}}
\put(94.00,11.00){\circle*{2.00}}
\put(94.00,40.00){\circle*{2.00}}
\put(79.00,11.00){\vector(1,0){14.00}}
\put(94.00,11.00){\vector(1,0){14.00}}
\put(109.00,40.00){\vector(-1,0){14.00}}
\put(94.00,40.00){\vector(-1,0){14.00}}
\put(94.00,5.00){\makebox(0,0)[cc]{$j+1$}}
\put(94.00,11.00){\line(0,1){29.00}}
\end{picture}
\caption[Integration contour on grid]{
   \label{HE_int_cont}
   Integration contour for the Heat Equation}
\end{center}
\end{figure}
Then, applying the midpoint rule for
the contour integral and the trapezoidal rule for the relation integral we
find two difference equations for two dependent variables $u, u_x$:
\begin{eqnarray}
  \left\{
   \begin{array} {l}
    \alpha \frac{\tau}{2}\, (1 + \theta_t - \theta_x^2 - \theta_t\theta_x^2) \circ u_x
    - 2\, h \,(\theta_x\theta_t - \theta_x) \circ u = 0 \\[0.2cm]
    \frac{h}{2}\, (\theta_x + 1) \circ u_x  - (\theta_x - 1) \circ u = 0\,.
   \end{array}
 \right. \label{DS for HE}
\end{eqnarray}

Furthermore, we show how to generate a finite-difference scheme for the Heat equation~(\ref{int_HE})
by using the Maple package LDA~\cite{GR'06}:

\medskip
\begin{maplegroup}
\begin{mapleinput}
\mapleinline{active}{1d}{with(LDA):}{%
}
\end{mapleinput}
\end{maplegroup}
\medskip
\noindent

First, we enter the independent and the dependent variables
for the problem.

\medskip

\begin{maplegroup}
\begin{mapleinput}
\mapleinline{active}{1d}{ivar := [j,k]; dvar := [ux,u]:}{%
}
\end{mapleinput}
\end{maplegroup}

\medskip
\noindent
Second, we translate (\ref{DS for HE}) into the input format
of the main command {\tt JanetBasis} in the package.

\medskip

\begin{maplegroup}
\begin{mapleinput}
\mapleinline{active}{1d}{L:=[a*t/2*(ux(j,k)+ux(j+1,k)-ux(j,k+2)-ux(j+1,k+2))-2*h*(u(j+1,k+1)-u(j,k+1)),
h/2*(ux(j,k+1)+u(j,k)-u(j,k+1)+u(j,k))]:}{%
}
\end{mapleinput}
\end{maplegroup}

\medskip

\noindent
Third, we compute the (minimal) Janet basis for $L$ w.r.t. an elimination ranking with $u_x\succ u$ to eliminate
the partial derivative $u_x$ from the system (\ref{DS for HE}). This ranking is chosen by using option $2$ as below;
in so doing we output only the element in Janet basis which does not contain $u_x$.

\medskip

\begin{maplegroup}
\begin{mapleinput}
\mapleinline{active}{1d}{JanetBasis(L,ivar,dvar,2)[1][1];}{%
}
\end{mapleinput}

\mapleresult
\begin{maplelatex}
\mapleinline{inert}{2d}{-2*a*t*u(j,k+1)+h*a*t*u(j,k)+2*a*t*u(j,k)+2*a*t*u(j,k+3)-h*a*t*u(j,k+2)-2*a*t*u(j,k+2)+2*a*t*u(j+1,k+3)-h*a*t*u(j+1,k+2)-2*a*t*u(j+1,k+2)+4*h^2*u(j+1,k+2)-4*h^2*u(j,k+2)-2*a*t*u(j+1,k+1)+h*a*t*u(j+1,k)+2*a*t*u(j+1,k);}{%
\maplemultiline{
 -\,2\,a\,t\,\mathrm{u}(j\,,k + 1) + h\,a\,t\,\mathrm{u}(j\,,k) + 2\,a\,t\,\mathrm{u}(j\,,k) + 2\,a\,t\,\mathrm{u}(j\,,k + 3) -
  h\,a\,t\,\mathrm{u}(j\,,k + 2)-2\,a\,t\,\mathrm{u}(j\,,k + 2)\\
 +\,2\,a\,t\,\mathrm{u}(j+1,k+3)-h\,a\,t\,\mathrm{u}(j+1,k+2)-2\,a\,t\,\mathrm{u}(j+1,k+2)+4\,h^2\,\mathrm{u}(j+1,k+2)\\-\,4\,h^2\,\mathrm{u}(j\,,k+2)
 -2\,a\,t\,\mathrm{u}(j+1,k+1)+h\,a\,t\,\mathrm{u}(j+1,k)+2\,a\,t\,\mathrm{u}(j+1,k) }
}
\end{maplelatex}
\end{maplegroup}

\medskip
\noindent
Thereby, we obtain the classical Crank-Nicolson scheme
\begin{eqnarray*}
 \frac{u_{k}^{j+1} - u_{k}^{j}}{ \tau}
 +  \alpha
\frac{(u_{k+1}^{j+1} - 2\, u_{k}^{j+1} + u_{k-1}^{j+1})
  +(u_{k+1}^{j} - 2\, u_{k}^{j} + u_{k-1}^{j})}{2\, h^{\,2}}=0\,,
\end{eqnarray*}
if in the above Maple output one shifts the second index by -1 and uses the first index as a superscript. The same
scheme is also obtained for the midpoint integration method applied to~(\ref{IR_for_HE}).

\section{Reduction of Feynman Integrals}

Evaluation of Feynman integrals is the cornerstone step of perturbative computations in elementary
particle physics~\cite{Smirnov'04}.
Consider, for example, a typical scalar  $L-$loop integral with $E$ external legs:
\begin{equation}
{\cal{I}}(\nu_1,\ldots,\nu_n)=\int d^dk_1\cdots d^dk_L \frac{1}{\prod_{j=1}^n D_j^{\nu_j}} \label{integral}
\end{equation}
which corresponds to $n$ internal lines in the related Feynman
diagram. Integration for every loop momentum $k_i$ is done over the space-time of
dimension $d=4-\epsilon$ where $\epsilon$ is the parameter of dimensional regularization~\cite{HW'74}.
The denominator
$D_j$ for the $j$-th internal line with mass $m_j$ is given by
$D_j:=p_j^2-m_j^2$. Here the line momenta $p_j$ are
linearly expressed in terms of the loop momenta $k_i$ $(i=1,\ldots,L)$ and external
momenta $q_s$ $(s=1,\ldots,E)$ as
$$ p_{j}=\sum_{s=1}^L \alpha_{js}k_{s}+\sum_{t=1}^E \beta_{jt}q_{t}\,,\qquad \alpha_{js},\beta_{jt}\in \Q\,.$$
Consider the combined set of $L+E$ vectors
$$r_a:=\left\{
\begin{array}{l}
k_a,\quad\ \ \ a=1,\ldots,L\,, \\
q_{a-L},\quad a=L+1,\ldots,L+E\,.
\end{array}
\right.
$$
Recurrence relations for integral (\ref{integral}) are derived by the integration-by-parts method~\cite{ChTk'81}
whose main idea is to use the integral identities (cf.~\cite{Smirnov'04,SS'03})
\begin{equation}
\int d^dk_1\cdots d^dk_L \frac{\partial}{\partial k_{i}}\cdot \frac{r_{j}}{\prod_{k=1}^n D_k^{\nu_k}}=0\, \label{IBP}
\end{equation}
together with the $d-$vector identities
\begin{equation}
2p_iq_j=(p_i+q_j)^2-(p_i^2-m_i^2)-(q_j^2+m_i^2)\,. \label{product}
\end{equation}
Integral identities~(\ref{IBP}) follow from an observation that any integral of $\partial /\partial k_i (\ldots)$
vanishes since there are no surface terms in dimensional regularization~(cf.~\cite{Gr'04}).

\medskip
\begin{figure}[ht]
\begin{center}
\begin{picture}(64,42)
\put(32,21){\makebox(0,0){\includegraphics{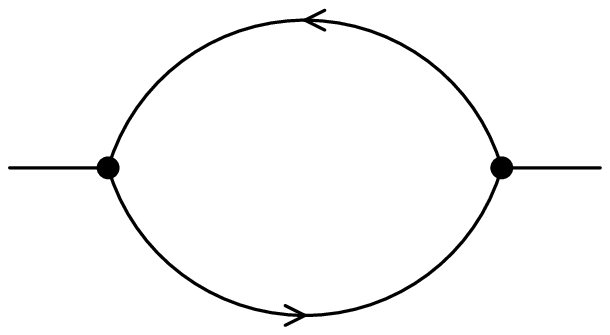}}}\put(6,24){\makebox(0,0)[t]{{q}}} \put(32,0){\makebox(0,0)[b]{{$k+q$}}}
\put(32,42){\makebox(0,0)[t]{{$k$}}} \put(57,24){\makebox(0,0)[t]{{q}}}
\put(32,8){\makebox(0,0)[b]{{$\nu_1$}}} \put(32,34){\makebox(0,0)[t]{{$\nu_2$}}}
\end{picture}
\end{center}
\caption{One-loop propagator diagram} \label{F:Q1}
\end{figure}
\medskip

\noindent
As a simple example consider one-loop propagator diagram of Fig.~\ref{F:Q1} with external momentum $q$ and with
one massive and another massless internal lines. This example was studied, for example,
in~\cite{Smirnov'04,Tarasov'98} and used already in~\cite{GR'06}. The corresponding Feynman integral~(\ref{integral})
is given by
\begin{equation}
{\cal{I}}(\nu_1,\nu_2)=\int  \frac{d^dk }{D_1^{\nu_1}D_2^{\nu_2}}\,,\qquad D_1:=(k+q)^2,\quad D_2:=k^2\,. \label{1-loop integral}
\end{equation}
In accordance to~(\ref{IBP}), there are two independent integral identities. Their integrands are
\begin{eqnarray*}
&& \frac{\partial}{\partial k}\cdot \frac{k}{D_1^{\nu_1}D_2^{\nu_2}}=\frac{1}{D_1^{\nu_1}D_2^{\nu_2}}\,\bigg[\frac{1}{\partial k}\cdot k-\frac{2\nu_1(k^2+q\cdot k)}{D_1}-
\frac{2\nu_2 k^2}{D_2}\bigg]\,,\\
&& \frac{\partial}{\partial k}\cdot \frac{q}{D_1^{\nu_1}D_2^{\nu_2}}=\frac{1}{D_1^{\nu_1}D_2^{\nu_2}}\,\,q\cdot\bigg[-\frac{2\nu_1(k+q)}{D_1}-
\frac{2\nu_2 k}{D_2}\bigg]\,.
\end{eqnarray*}
Taking into account the identity $2k\cdot q=(k+q)^2-(k^2-m^2)-(q^2+m^2)$ of type~(\ref{product})
and equality $\partial /\partial k\cdot k=d$ we obtain the difference system
\begin{equation}
\left\{
\begin{array}{l}
\big[d-\nu_1- 2\nu_2-\nu_1\theta_1\theta_2^{(-1)}+\nu_1(q^2-m^2)\theta_1-2m^2\nu_2\theta_2\big]\circ {\cal{I}}(\nu_1,\nu_2)=0\,,\\
\big[\nu_1-\nu_2+\nu_1(q^2-m^2)\theta_1-\nu_1\theta_1\theta_2^{(-1)}+\nu_2\theta_1^{(-1)}\theta_2-\nu_2(q^2+m^2)\theta_2\big]\circ
{\cal{I}}(\nu_1,\nu_2)=0\,, \label{dsystem}
\end{array}
\right.
\end{equation}
where $\theta_i^{(-1)}$ and $\theta_2^{(-1)}$ denote the left-shift operators for indices $\nu_1$ and $\nu_2$,
respectively.

Now we construct the minimal set of master or basic integrals for the two-indexed family~(\ref{1-loop integral}) of
Feynman integrals by applying the Maple package LDA~\cite{GR'06} with the input denotations
$k:=\nu_1,n:=\nu_2$ and $f(k+1,n+1):={\cal{I}}(\nu_1,\nu_2)$:

\medskip

\begin{maplegroup}
\begin{mapleinput}
\mapleinline{active}{1d}{ivar:=[k,n]: dvar:=[f]:}{%
}
\end{mapleinput}
\end{maplegroup}

\medskip

Then, we enter the recurrence relations (\ref{dsystem}).
\medskip

\begin{maplegroup}
\begin{mapleinput}
\mapleinline{active}{1d}{L:=[(d-k-2*n)*f(k+1,n+1)-k*f(k+2,n)+k*(q^2-m^2)*f(k+2,n+1)-2*m^2*n*f(k+1,n+2),
(k-n)*f(k+1,n+1)+k*(q^2-m^2)*f(k+2,n+1)-k*f(k+2,n)+n*f(k,n+2)-
n*(q^2+m^2)*f(k+1,n+2)]:}{%
}
\end{mapleinput}
\end{maplegroup}
\medskip
As the next step we compute a Janet basis for an orderly ranking (Sect.2) induced by $\theta_1\succ \theta_2$.
\medskip
\begin{maplegroup}
\begin{mapleinput}
\mapleinline{active}{1d}{JB:=JanetBasis(L,ivar,dvar):}{%
}
\end{mapleinput}
\end{maplegroup}

\medskip
In order to compute the set of master integrals we have to take into account that ${\cal{I}}(\nu_1,\nu_2)=0$
for $\nu_2\leq 0$~\cite{Smirnov'04,Gr'04}. This extra {\it boundary} information is input as

\medskip

\begin{maplegroup}
\begin{mapleinput}
\mapleinline{active}{1d}{AddRelation(f(k+j,n)=0,ivar,dvar):}{%
}
\end{mapleinput}
\end{maplegroup}

\medskip
Master integrals are those $f(k,n)$ which are independent modulo all the consequences $R$ (see Sect.2)
of~(\ref{dsystem}). Thereby, the master integrals are easily determined via the leading terms of the Janet basis. Namely,
one has to determine those $f(k,n)$ that are not expressible as the action of a power product
$\theta_1^{\mu_1}\theta_2^{\mu_2}$ $(\mu_1,\mu_2\in \N_{\geq 0})$ on a leading term in the
Janet basis (cf. definition (\ref{GB})).

The set of master integrals is computed by invoking the command:

\medskip
\begin{maplegroup}
\begin{mapleinput}
\mapleinline{active}{1d}{ResidueClassBasis(ivar,dvar);}{%
}
\end{mapleinput}

\mapleresult
\begin{maplelatex}
\mapleinline{inert}{2d}{[f(k,n+1), f(k,n+2), f(k+1,n+1)];}{%
\[
[\mathrm{f}(k, \,n + 1), \,\mathrm{f}(k, \,n + 2), \,
\mathrm{f}(k + 1, \,n + 1)]
\]
}
\end{maplelatex}
\end{maplegroup}

\medskip

Now any integral $f(k+i,n+j)$ can be explicitly expressed as a linear combination of the master integrals whose
coefficients are rational functions
in parameters $d,q^2$ and $m^2$. The explicit expression is obtained algorithmically by applying the \Gr or Janet
reductions described in Sect.2.
In LDA the Janet reductions are performed. To show the output of such an expression for $f(k+3,n+2)$ and make the
output more compact
we let $m=0$ and show the underlying piece of the Maple code:

\medskip

\begin{maplegroup}
\begin{mapleinput}
\mapleinline{active}{1d}{m:=0: J:=JanetBasis(L,ivar,dvar):}{%
}
\end{mapleinput}
\end{maplegroup}
\begin{maplegroup}
\begin{mapleinput}
\mapleinline{active}{1d}{AddRelation(f(k,n+j)=0,ivar,dvar):}{%
}
\end{mapleinput}
\end{maplegroup}
\begin{maplegroup}
\begin{mapleinput}
\mapleinline{active}{1d}{ResidueClassBasis(ivar,dvar);}{%
}
\end{mapleinput}

\mapleresult
\begin{maplelatex}
\mapleinline{inert}{2d}{[f(k+1,n+1)];}{%
\[
[\mathrm{f}(k + 1, \,n + 1)]
\]
}
\end{maplelatex}
\end{maplegroup}
\begin{maplegroup}
\begin{mapleinput}
\mapleinline{active}{1d}{InvReduce(f(k+3,n+2),J,"F");}{%
}
\end{mapleinput}

\mapleresult
\begin{maplelatex}
\mapleinline{inert}{2d}{-(2*n+4-d+2*k)*(2*n+2-d+2*k)*(2*k+n-d)*(n+3-d+k)*(n+2-d+k)*f(k+1,n+1
)/((n+1)*(2*n-d+4)*n*q^6*k*(d-2*k-2));}{%
\maplemultiline{
 - ((d - 2 - 2\,k - 2\,n)\,( d - 4 - 2\,k - 2\,n)(d - 2 - k - n)\,(d - 3 - k - n)\,(- n - 2\,k + d )\,\mathrm{f}(k + 1, \,n + 1))/ \\
( q^{6}\,(k + 1)\,(-2\,k + d - 4)\,k \,(-2\,k - 2 + d)\,n) }
}
\end{maplelatex}
\end{maplegroup}

\medskip

In the massless case $(m=0)$ a new extra relation $f(k,n+j)$ equivalent to the {\it boundary condition}
${\cal{I}}(\nu_1,\nu_2)=0$ for $\nu_1\leq 0$ is added that yields the only master integral $f(k+1,n+1)$. The last command invokes the LDA procedure \ that computes
the $J-$normal form of $f(k+3,n+2)$ modulo the Janet basis. This normal form just represents $f(k+3,n+2)$ in terms of the master integral.
Option ''F'' provides factorization of the numerator and denominator in the output rational function coefficient. It should be noted
that, since integral ${\cal{I}}(\nu_1,\nu_2)$ is non-vanishing only when both its indices are positive, the master integral can be identified
with ${\cal{I}}(1,1)$.

\section{Conclusion}

We shown above that the \Gr bases technique can be applied to generate difference schemes for linear PDEs and to reduce
multiloop Feynman integrals. Each of our simple illustrative examples of Sect.3 and 4 needs less than 1 second of computing time
on an 1.7 Mhz personal computer with 512 Mb RAM. Larger examples, however, can require much more computer resources since complexity
of a \Gr basis computation is at least singly exponential, and may be even doubly exponential, in a number of
variables~\cite{BW'98,zuGathen'03}.
Besides, blowing-up of intermediate coefficients, especially in the presence of parameters, as in the case of recurrence relations for
Feynman integrals, is a serious obstacle in the practice. That is why to apply \Gr bases to multivariate and multiparametric problems
one has not only to optimize and improve the underlying algorithms and data structures but also to implement them in lower level languages than
Maple or Mathematica. Our Janet division algorithms~\cite{G'05} have already implemented in C and C++ (see the Web page~\cite{invo})
for commutative polynomials, and extension of these codes to differential and difference equations is planned for the coming years.
As it is argued in~\cite{Tarasov'98,Tarasov'04}, differential \Gr bases can also be applied to reduction of Feynman integrals.
A practical specialization of the \Gr bases ideas to reduction of Feynman integrals was suggested
recently in~\cite{SS'05,S'06} where the whole index space for integrals~(\ref{integral}) is partitioned into
so-called {\em sectors} in accordance to the extra {\it boundary conditions}. Then in every sector a certain kind of a
Gr\"{o}bner-like
basis is constructed. As to an extended discussion of generating difference schemes by means of \Gr bases we refer to our recent
paper~\cite{GBM'06}.

\section{Acknowledgements}
The research presented in the paper was partially supported by grants 04-01-00784 and
05-02-17645 from the Russian Foundation for Basic Research and by grant 5362.2006.2 from
the Ministry of Education and Science of the Russian Federation.

\end{document}